\documentclass[11pt]{iopart}

\begin{document}

\title{Three-geometry and reformulation of the Wheeler-DeWitt equation}

\author{Chopin Soo}

\address{Department of Physics, National Cheng Kung University,
Tainan 701, Taiwan}

\ead{cpsoo@mail.ncku.edu.tw}

\begin{abstract}
A reformulation of the Wheeler-DeWitt equation which highlights
the role of gauge-invariant three-geometry elements is presented.
It is noted that the classical super-Hamiltonian of
four-dimensional gravity as simplified by Ashtekar through the use
of gauge potential and densitized triad variables can furthermore
be succinctly expressed as a vanishing Poisson bracket involving
three-geometry elements. This is discussed in the general setting
of the Barbero extension of the theory with arbitrary
non-vanishing value of the Immirzi parameter, and when a
cosmological constant is also present. A proposed quantum
constraint of density weight two which is polynomial in the basic
conjugate variables is also demonstrated to correspond to a
precise simple ordering of the operators, and may thus help to
resolve the factor ordering ambiguity in the extrapolation from
classical to quantum gravity. Alternative expression of a density
weight one quantum constraint which may be more useful in the spin
network context is also discussed, but this constraint is
non-polynomial and is not motivated by factor ordering. The
article also highlights the fact that while the volume operator
has become a preeminient object in the current manifestation of
loop quantum gravity, the volume element and the Chern-Simons
functional can be of equal significance, and need not be mutually
exclusive. Both these fundamental objects appear explicitly in the
reformulation of the Wheeler-DeWitt constraint.

\end{abstract}



\maketitle

\newcommand*{\ket}[1]{\ensuremath{|#1\rangle}}
\newcommand{\tE}{\tilde E}
\newcommand{\tB}{\tilde B}
\newcommand*{\eps}{{\rlap{\lower2ex\hbox{$\,\,\tilde{}$}}{\epsilon_{ijk}}}}
\newcommand{\sig}{\tilde\sigma}
\newcommand{\ep}{\epsilon}
\newcommand{\ept}{\tilde\epsilon}
\newcommand{\be}{\begin{equation}}
\newcommand{\en}{\end{equation}}
\newcommand{\cK}{\cal K}
\newcommand{\tv}{\tilde v}

\section{Introductory remarks}


The program of non-perturbative canonical quantization of gravity
attempts to overcome the perturbative non-renormalizability of
Einstein's theory by constructing exact background-independent
quantum geometry. Much excitement and insight have stemmed from
Ashtekar's reformulation of the Hamiltonian theory and the
simplification of the constraints through the use of gauge
connection and densitized triad variables\cite{Ash}. Conceptually,
the distinction between geometrodynamics and gauge dynamics is
bridged by the identification of the densitized triad, $\tE^{ia}$
-from which the metric is a derived composite - as the momentum
conjugate to the gauge potential $A_{ia}$. Most intriguing too is
the conjunction of the fact the Lorentz group possesses self and
anti-self-dual decompositions in four and only in four dimensions
with the observation that the Ashtekar-Sen connection is precisely
the pullback to the Cauchy surface of the self-(or anti-self)-dual
projection of spin connection\cite{SJS}. The infusion of loop
variables\cite{RS-loop} and subsequently spin network
states\cite{RS-spinnetwork} have also proved fruitful, and have
yielded discrete spectra for well-defined area and volume
operators\cite{Area-Vol}. Indeed  by virtue of being area and
volume eigenstates, states based upon spin networks -the latter
originally introduced by Penrose to explore quantum
geometry\cite{Penrose}- are now prominent in the current
manifestation of loop quantum gravity. To the extent that exact
states and rigorous results are needed, simplifications of the
classical and corresponding quantum constraints are crucial steps
indeed. These include Ashtekar's original simplification as well
as Thiemann's observation that
$\epsilon_{abc}\eps\tE^{ia}\tE^{jb}$ in the super-Hamiltonian
constraint is proportional to the Poisson bracket between the
connection and the volume operator\cite{Thiemann}\footnote{$SO(3)$
indices are denoted by lower case Latin letters at the beginning
of the alphabet from a to h, while spatial indices on the
3-dimensional Cauchy surface are denoted by Latin letters from i
onwards.}.

In this article a reformulation of the super-Hamiltonian
constraint and its associated Wheeler-DeWitt Equation is
presented. It is noted that the classical super-Hamiltonian of
four-dimensional gravity as simplified by Ashtekar through the use
of gauge potential and densitized triad variables can furthermore
be succinctly expressed as a vanishing Poisson bracket between
fundamental invariants. This is discussed in the general setting
of the Barbero extension of the theory\cite{Barbero}, with
arbitrary non-vanishing value of the Immirzi parameter
$\gamma$\cite{Immirzi}, and when a cosmological constant $\lambda$
is also present. The observation naturally suggests a
reformulation of quantum gravity wherein the Wheeler-DeWitt
equation is reduced to the requirement of the vanishing of the
corresponding commutator. Alternative ways of expressing the
quantum constraint will also be discussed.

It has long been known that 3-dimensional diffeomorphism
invariance would require the quantum states to be functionals of
3-geometries\cite{Wheeler,DeWitt}. It may therefore be surmised
that, albeit a nontrivial endeavor, it ought to be possible to
express the Wheeler-DeWitt Equation of the full theory in terms of
explicit 3-geometry elements. However the constraint is also
required to be satisfied at each point on the Cauchy surface. Both
these requirements are remarkably realized in the reformulation
here in that the Wheeler-DeWitt constraint is equivalent to the
vanishing of the commutator between $\tv^2({\vec x}) =
[\det(\tE^{ia}({\vec x}))]$ at each point and a combination of the
integrals involving the extrinsic curvature and the Chern-Simons
functional of the gauge connection. The reformulation not only
highlights the role of gauge-invariant 3-geometry elements in the
Wheeler-DeWitt Equation, but also spells out which specific
superspace functionals are involved.

The Chern-Simons functional has served as a fertile link between
quantum field theories of three and four dimensions. In General
Relativity with Ashtekar variables, it has the additional
significance of being the carrier of information of both intrinsic
and extrinsic curvatures (see for instance Eq.(6) later). Although
it is not as extensively explored in spin networks in present
formulations of loop quantum gravity as in quantum field theories,
the expectation that the Chern-Simons invariant has a very
significant, and even direct, role in 4-dimensional quantum
gravity is in fact borne out by the discussions in this article.
It should also be emphasized that while the volume operator has
become a preeminient object in the current manifestation of loop
quantum gravity, the volume element and the Chern-Simons
functional can be of equal significance and need not be mutually
exclusive. Both these fundamental objects appear explicitly in the
reformulation of the Wheeler-DeWitt constraint.

There is another feature of the reformulation which is worth
emphasizing. Unlike the Gauss Law and super-momentum constraints
which are kinematic and have straightforward group-theoretic
interpretations, the factor ordering ambiguity of the
non-commuting variables in quantum super-Hamiltonian constraint is
a more intricate matter. There is no unique prescription for
defining a quantum theory from its classical correspondence. Thus
the factor ordering problem has to be decided through other means;
for instance, through mathematical consistency (sometimes
expediency) and the absence of quantum anomalies. Even so, these
may or may not yield a unique ordering. With complex Ashtekar
variables, Hermiticity of the super-Hamiltonian too cannot be
adopted as a criterion. Often when dealing with the factor
ordering of a complicated constraint, an initial motivation is
needed; and a specific ordering is assumed first before checking
the consistency of the composite operator.

A proposed quantum constraint of density weight two motivated by
3-geometry considerations here and which is also polynomial in the
basic conjugate variables will be demonstrated to correspond to a
precise simple ordering of the quantum operators, and may thus
help to resolve the factor ordering ambiguity in the extrapolation
from classical to quantum gravity. However, it has also been
pointed out background independent field theories are ultraviolet
self-regulating if the constraint weight is equal to one but not
for other density weights\cite{Thiemann}. To wit, we also discuss
an alternative density weight one quantum constraint which may be
more useful in the spin network context; but this expression is
non-polynomial in the basic conjugate variables, and it is not
motivated by factor ordering.

\section{Reformulation of the classical super-Hamiltonian constraint}

Starting with the fundamental conjugate pair and Poisson bracket,
\be \{{\tE}^{ia}({\vec x}),k_{bj}({\vec y})\}_{P.B.} =
\beta\delta^i_j \delta^a_b \delta^3({\vec x} - {\vec y}),
\label{e:abb}\en with $\beta \equiv (\frac{8\pi G}{c^3}) =
\frac{8\pi l^2_p}{\hbar}$ (where $l_p$ is the Planck length), the
Barbero extension\cite{Barbero} of $k_{ia} = E^{j}_aK_{ij}$ to a
generalized Ashtekar $SO(3)$ gauge connection\footnote{It has been
pointed out that this generalized connection is however not the
pullback onto spatial slices of a four-dimensional spin connection
unless $\gamma =\pm i$\cite{Samuel}.}, \be A^\gamma_{ai} \equiv
\gamma k_{ai} + \Gamma_{ai}, \en yields
 \be \{{\tE}^{ia}({\vec x}),A^\gamma_{bj}({\vec y})\}_{P.B.} =
\gamma\beta\delta^i_j \delta^a_b \delta^3({\vec x} - {\vec
y}).\label{e:abc}\en In the above $\Gamma_a$ is the torsionless
($dE_a + \epsilon_{ab}\,^c\Gamma_b \wedge E_c = 0$) connection
compatible with the dreibein 1-form $E_a= E_{ai}dx^i$ on the
Cauchy manifold $M$; and $K_{ij}$ denotes the extrinsic curvature.
In terms of 3-dimensional torsionless spin connection
$\omega_{ab}$, $\Gamma_{ai} =
-\frac{1}{2}\epsilon_a\,^{bc}\omega_{ibc}$.

Gauss Law constraint for $SO(3)$ gauge invariance is equivalent to
\be 0 \approx \epsilon^{ab}\,_ck_{ib}\tE^{ic} =
\frac{1}{\gamma}D^A_i \tE^{ia}. \en  $D^A$ means the covariant
 derivative  with respect to the connection $A$; when there is no danger of
confusion, we shall suppress the $\gamma$ index in the connection
and denote it simply by $A_{ai}$, with implicit
 dependence on the Immirzi parameter $\gamma$.

Four-dimensional General Relativity as a theory of the conjugate
pair of densitized triad and gauge variables, $(\tE^{ia},
A_{ai})$, seems to be anchored on a few fundamental physical
objects: the volume element $(\tv)$, the Chern-Simons functional
of the gauge potential (${\cal C}[A]$), and invariants constructed
from the extrinsic curvature (${\cK}$ and ${\cal D}$). All are
gauge invariant, but apart from $\tv$ which is a tensor density,
they are in addition also invariant under three-dimensional
diffeomorphisms i.e. they are elements of 3-geometry. The
definitions for these objects will be discussed below:
\be\tv({\vec x}) \equiv
\sqrt{\frac{1}{3!}\epsilon_{abc}\eps\tE^{ia}({\vec
x})\tE^{jb}({\vec x})\tE^{kc}({\vec x})} = |\det E_{ai}|.
\label{e:m}\en Its integral over the Cauchy surface, $M$, is the
volume, $V =\int_M \tv{({\vec x}}) d^3x$. The Chern-Simons
functional is
\begin{eqnarray} {\cal C}[A^\gamma] &\equiv& \frac{1}{2}\int_M (A^a \wedge dA_a +
\frac{1}{3}\ep^{abc}A_a\wedge A_b\wedge A_c) \cr &=& {\cal
C}[\Gamma] +\gamma \int_M R_\Gamma^a\wedge k_a  +
\frac{\gamma^2}{2!}\int_M k^a \wedge (D^\Gamma k)_a \cr &&+
\frac{\gamma^3}{3!} \int_M \epsilon^{abc}k_a \wedge k_b \wedge k_c
\label{e:kkk}.
\end{eqnarray}\footnote{Its characteristic feature is that it satisfies
$\frac{{\delta}{\cal C}[A]}{\delta{A_{ia}}}={\tilde B}^{ia}$ if
$\partial M = 0$; wherein $\tB^{ia}$ is the non-Abelian $SO(3)$
magnetic field of $A_{ai}$. If $M$ is with boundary, additional
considerations need to be taken into account e.g. the imposition
of appropriate boundary conditions, or consideration of whether
the addition of a supplementary boundary term to ${\cal C}$ can
again render the super-Hamiltonian to be a Poisson bracket. On the
other hand, one can treat $\partial M =0$ as a predictive element
of the formulation discussed here.}. In the last equality we have
expanded the Chern-Simons functional for $A^\gamma = \gamma k +
\Gamma$ in terms of ${\cal C}[\Gamma]$ which is the Chern-Simons
functional for the connection $\Gamma$, and higher order terms;
and $R_\Gamma$ is the curvature 2-form of the connection $\Gamma$.

The definition of $\Gamma$ also implies the integral of the trace
of the extrinsic curvature can be expressed in a couple of ways:
\be{\cK} \equiv \frac{1}{2\gamma}\int_M E^a \wedge (D^{A} E)_a =
\int_M ( {\tE}^{ia}k_{ai} ) d^3x.\label{e:kkl}\en Moreover,
${\cK}$ can be written totally in terms of the volume $V$ and $A$
from the observation that the dreibein $E_{ai}$ (inverse of the
triad $E^{ia}$) can also be expressed as \be E_{ai} =
\frac{2}{\gamma\beta} \{V, A_{ai}\}_{P.B.} .\label{e:pc}\en

With the above definitions and the fundamental relation of
Eq.({\ref{e:abc}}), it follows that the Poisson brackets below are
true:
\begin{eqnarray}
\{{\tE}^{ia}({\vec x}), {\cal C}[A^\gamma] \}_{P.B.}&=&
(\beta\gamma)\tB^{ia}_\gamma({\vec x})\cr \{{\tE}^{ia}({\vec x}),
{\cK} \}_{P.B.}&=& \beta\tE^{ia}({\vec x})\cr
 \{ \tv^2, {\cK}
\}_{P.B.}&=& {3\beta}\tv^2
\cr \{\tv^2, {\cal C}[A^\gamma] \}_{P.B.} &=&
\frac{\beta\gamma}{2}{\epsilon_{abc}\eps\tE^{ia}\tE^{jb}\tB^{kc}_\gamma}\cr
\{{\cK}, {\cal C}[A^\gamma] \}_{P.B.}&=& \beta\int_M
(\tB^{ia}_\gamma k_{ai})\, d^3x\cr \{ \tv^2, \frac{{\cal
C}[A^\gamma]}{\gamma} + \frac{\lambda}{3}{\cK}
\}_{P.B.}&=&\frac{\beta}{2}\epsilon_{abc}\eps\tE^{ia}\tE^{jb}(\tB^{kc}_\gamma
+ \frac{\lambda}{3}\tE^{kc}).\label{e:pb}
\end{eqnarray}
$\tB^{{ia}^{\gamma}} =
\frac{1}{2}\tilde\epsilon^{ijk}F^{{a}^{\gamma}}_{jk}$ is the
magnetic field of $A^\gamma$; with \be F^{\gamma}_{aij} =
R^\Gamma_{aij} + \gamma(D^\Gamma_ik_{aj} -D^\Gamma_jk_{ai}) +
\gamma^2 \epsilon^{abc}k_{bi}k_{cj}.\en

Following Ref.\cite{Barbero}, the usual ADM super-Hamiltonian
constraint $\tilde H \approx 0 $ can be rewritten with
 \begin{eqnarray}
&&\tv\tilde H \propto (\tv^2)[Tr(K^2) - (Tr K)^2 - R^\omega +
2\lambda]\cr &=&
\epsilon_{abc}\eps\tE^{ia}\tE^{jb}[\tB^{kc}_\gamma
-\frac{(1+\gamma^2)}{2}\epsilon^{cde}\tilde\epsilon^{klm}k_{dl}k_{em}
+ \frac{\lambda}{3}\tE^{kc}].\label{e:pt}\end{eqnarray} $R^\omega$
is just the Ricci scalar curvature of the spin connection
$\omega$. In the above equality we have used
$\epsilon_{ab}\,^c\tE^{ia}\tE^{jb}D^\omega_ik_{cj}$$ \propto
\ept^{ijk}\tE^a_{i}(D^\Gamma_j k_{k})_a$$ = 0$ by virtue of
$\ept^{ijk}D^\Gamma_j\tE_{ak}=0$ and the Gauss Law constraint
which implies $\ept^{ijk}\tE_{aj}k^a_k =0$.

We may introduce another gauge-invariant 3-geometry element
(essentially the integral of determinant of $k_{ia}$):
 \be {\cal D} \equiv \frac{1}{3!}\int_M\epsilon^{abc}k_a\wedge
 k_b\wedge k_c = \frac{1}{3!}\int_M\ept^{ijk}\epsilon^{abc}k_{ai}
 k_{bj}k_{ck} d^3x \en
which has the properties \begin{eqnarray} \{ \tE^{ia}, {\cal
D}\}_{P.B.} &=&
 \frac{\beta}{2}\epsilon^{abc}\ept^{ijk}k_{bj}k_{kc}\cr
 \{\tv^{2}, {\cal D}\}_{P.B.} &=&
 \frac{\beta}{4}\epsilon_{abc}\eps\tE^{ia}\tE^{jb}\epsilon^{cde}\ept^{klm}k_{dl}k_{em}.
 \end{eqnarray}
Using this last identity and the final Poisson bracket in
({\ref{e:pb}}), we can verify that it is possible to {\it express
the complete super-Hamiltonian constraint as a vanishing Poisson
bracket} \be 0 \approx \tv \tilde H \propto \{ \tv^2, \frac{{\cal
C}[A^\gamma]}{\gamma} + \frac{\lambda}{3}{\cal K}
-(1+\gamma^2){\cal D}\}_{P.B.}\en The self-dual and antiself-dual
specializations correspond to $\gamma=\mp i$ which simplify the
expression by eliminating ${\cal D}$ which is cubic in $k$.
Szabados first noted that the super-Hamiltonian constraint is
expressible as a Poisson bracket for the further specialization of
vanishing cosmological constant\cite{Szabados}.

\section{Quantization and reformulation of the Wheeler-DeWitt equation}

Even though we may invoke Poisson bracket-quantum commutator
correspondence $\{ ,\}_{P.B.}\mapsto (i\hbar)^{-1}[\,,\,] $, there
is no unique prescription for defining a quantum theory from its
classical correspondence. The previous observations naturally
suggest defining four-dimensional non-perturbative quantum General
Relativity as a theory of the conjugate pair $(\tE^{ia}, A_{ai})$
with super-Hamiltonian constraint imposed as \be [\tv^2,
\frac{{\cal C}[A^\gamma]}{\gamma} + \frac{\lambda}{3}{\cal K}
-(1+\gamma^2){\cal D}] =0; \label{e:abf}\en together with the
requirement of invariance of the theory under three-dimensional
diffeomorphisms and $SO(3)$ gauge transformations of $(\tE, A)$.

This reformulation is bolstered by the existence of a precise
factor ordering of the non-commuting operators which realizes the
quantum Wheeler-DeWitt constraint. To check this we may utilize
repeatedly for composite operators the commutator identities
$[XY,Z]= X[Y,Z] + [X,Z]Y$ and $[X,YZ] = [X,Y]Z + Y[X,Z]$. Thus \be
\fl
[\tv^2=\frac{1}{3!}\epsilon_{abc}\eps\hat{\tE^{ia}}\hat{\tE^{jb}}\hat{\tE^{kc}},
{\cal C}[A] ] = (8\pi{i}{l^2_p}\gamma)\frac{1}{3!}
\eps\epsilon_{abc}(\hat{\tE^{ia}}\hat{\tE^{jb}}\hat{\tB^{kc}} +
\hat{\tE^{ia}}\hat{\tB^{jb}}\hat{\tE^{kc}} +
\hat{\tB^{ia}}\hat{\tE^{jb}}\hat{\tE^{kc}}),\en if we take into
account $[\hat{\tE^{ia}}, \hat{\cal C}[A] ] =
(8\pi{i}{l^2_p}\gamma)\tB^{ia}$ which follows from the fundamental
commutation relation $[\hat{\tE^{ia}}({\vec x}),
\hat{A_{bj}}({\vec y}) ] = 8\pi{i}\gamma{l^2_p}\delta^i_j
\delta^a_b \delta^3({\vec x} - {\vec y})\ $. In a similar manner,
\be\fl [\tv^2, {\cal D}] =
(8\pi{i}{l^2_p}\gamma)\frac{1}{3!}\epsilon^{abc}\eps\epsilon_c\,^{de}
\frac{1}{2}\ept^{klm}(\hat{\tE^{ia}}\hat{\tE^{jb}}\hat{k_{dl}}\hat{k_{em}}
+ \hat{\tE^{ia}}\hat{k_{dl}}\hat{k_{em}}\hat{\tE^{jb}} +
\hat{k_{dl}}\hat{k_{em}}\hat{\tE^{ia}}\hat{\tE^{jb}}). \en Thus
Eq.({\ref{e:abf}}) corresponds to the precise factor ordering of
the super-Hamiltonian constraint which is \begin{eqnarray}
&&\tilde{\tilde H} \propto \epsilon_{abc}\eps(
\frac{1}{3}(\hat{\tE^{ia}}\hat{\tE^{jb}}\hat{\tB^{kc}} +
\hat{\tE^{ia}}\hat{\tB^{jb}}\hat{\tE^{kc}} +
\hat{\tB^{ia}}\hat{\tE^{jb}}\hat{\tE^{kc}}) +
\frac{\lambda}{3}\hat{\tE^{ia}}\hat{\tE^{jb}}\hat{\tE^{kc}}\cr
&&-\frac{(1+\gamma^2)}{2}\frac{1}{3}
\epsilon^{cde}\tilde\epsilon^{klm}(\hat{\tE^{ia}}\hat{\tE^{jb}}\hat{k_{dl}}\hat{k_{em}}
+ \hat{\tE^{ia}}\hat{k_{dl}}\hat{k_{em}}\hat{\tE^{jb}} +
\hat{k_{dl}}\hat{k_{em}}\hat{\tE^{ia}}\hat{\tE^{jb}})).
\label{e:abg} \end{eqnarray} From the general quantum
Wheeler-DeWitt constraint of Eq.(\ref{e:abf}) the original
Ashtekar self-dual and antiself-dual specializations (with $\gamma
= \mp i$) may therefore be expressed succinctly as \be
[\hat{\tv^2}({\vec x}), \hat{\cal C}[A] ] =
\pm\lambda(8\pi{l^2_p})\hat{\tv^2}({\vec x}).\label{e:abl}\en This
corresponds to the {\it symmetric ordering}, wherein the operators
$\tE$ and ${\tilde B}$ appear in every permutation in the
constraint with equal weight for each combination(as indicated in
Eq.({\ref{e:abg}}) with $\gamma =\mp i$). In the classical limit
with commuting operators, (\ref{e:abg}) reduces to (\ref{e:pt}).
Note also that although it is possible to express $\tv^2$ on the
R.H.S. of Eq.(\ref{e:abl}) as the commutator
$\frac{1}{3(8\pi{l^2_p}i)}[\tv^2,{\cal K}]$ \footnote{The physical
meaning of the operator $\hat{\cK}$ is not readily apparent, but
we may nevertheless deduce its connection to ``intrinsic time" in
quantum gravity from the Poisson bracket $\{\tv^2(x),
{\cK}\}_{P.B}= 3\beta\tv^2(x)$; so apart from a multiplicative
constant, ${\cK}$ is in fact conjugate to the variable $\ln\tv$.
In the quantum context $\hat{\cK}$ is thus proportional to the
{\it generator of translations} in $\ln\tv =
\ln|\det{E_{ai}}|$\cite{soocp}. This variable is furthermore a
monotonic function of the superspace ``intrinsic time variable"
($\propto \sqrt{|\det E_{ai}|}$) discovered by DeWitt in his
seminal study of canonical quantum gravity and the signature of
the supermetric many years ago\cite{DeWitt}.}, it may not be
expedient to do so (we shall discuss the related issues shortly)
if we do not insist on writing the constraint as a vanishing
commutator relation. With Dirac quantization, the reformulated
Wheeler-DeWitt Equation is therefore\footnote{Ref.\cite{soocp}
contains related discussions on the reformulation.} \be
[\hat{\tv^2},{\hat C}[A] ] \ket{\Psi_{Phys}} =
\pm\lambda(8\pi{l^2_p})\hat{\tv^2}\ket{\Psi_{Phys}}.\label{e:WD}\en
It has a number of remarkable properties:(1)This equation for the
{\it full theory} (not just a particular minisuperspace sector) is
{\it not merely symbolic} but is in fact {\it expressed
explicitly} in terms of the gauge-invariant 3-geometry element
which is none other than the Chern-Simons functional of the
Ashtekar connection. This is to be contrasted with the traditional
formulation with metric variables, wherein the Equation was {\it
symbolically}\cite{Wheeler} \be[\frac{\delta^2}{{\delta{\cal
G}}^2} +(R({\cal G})-2\lambda)]\ket{\Psi_{Phys.}} =0.\en (2)There
are now {\it no factor ordering ambiguities} for the composite
operators $\tv^2 =
\frac{1}{3!}\epsilon_{abc}\eps\tE^{ia}\tE^{jb}\tE^{kc}$ and ${\cal
C}[A] =   \frac{1}{2}\int_M (A^a \wedge dA_a +
\frac{1}{3}\ep^{abc}A_a\wedge A_b\wedge A_c)$ which have clear
geometric meanings; and each of which is made up of {\it
commuting} variables (this too is true of the operator ${\cal D}$
when $\gamma \neq \mp i$ is adopted in the more general context of
Eq.(\ref{e:abf})). Whatever ordering ``ambiguities" that were
present in super-Hamiltonian constraint in the transition from
classical to quantum theory have been decided by requiring that
the Wheeler-DeWitt Equation is expressible in terms of these
3-geometry elements. (3)The operators $\tv^2$, ${\cal C}[A]$, and
also the combination ${\lambda l^2_P}$, which appear in the
reformulated Wheeler-DeWitt Equation are now all {\it
dimensionless}.

So far the reformulation is not confined to specific
representations in the quantum theory, and we have used $\tv^2$,
which is polynomial in $\tE$, instead of $\tv$ to express the
constraint in polynomial form. This results in a super-Hamiltonian
of density weight two. It has been pointed out background
independent field theories are ultraviolet self-regulating if the
constraint weight is equal to one but not for other density
weights\cite{Thiemann}. Thus it may be desirable to use $\tv$
instead and conjecture an alternative Wheeler-DeWitt equation of
the form
 \be [\, {\hat{\tv}},{\hat{\cal C}}[A] \,] \ket{\Psi_{Phys}} =
\pm\lambda(4\pi{l^2_p}){\hat{\tv}}\ket{\Psi_{Phys}}.\label{e:WD}\en
Although $\tv$ and the associated density weight one constraint is
non-polynomial in the basic variables, this version of the
constraint does lead to a necessary condition involving the volume
operator ${\hat V} =\int \hat{\tv} \,d^3x$ that is \be [\, {\hat
V},{\hat{\cal C}}[A] \,] \ket{\Psi_{Phys}} =
\pm\lambda(4\pi{l^2_p}){\hat V}\ket{\Psi_{Phys}}.\label{e:WE}\en
Explicit realizations and representations of eigenstates of the
volume operator can be precisely associated with spin network
states\cite{Area-Vol}. Furthermore the volume operator is
classically real and in the quantum context should be Hermitian,
implying its eigenstates form a complete basis; and all physical
states can be expanded in terms of these spin network volume
eigenstates.  Instead of treating $\tv$ as
$\sqrt{\frac{1}{3!}\epsilon_{abc}\eps\tE^{ia}\tE^{jb}\tE^{kc}}$,
an alternative expression in terms of the dreibein $E_{ai}$
(inverse to the triad) which is also of interest is $\tv =
\frac{1}{3!}{\tilde\epsilon}^{ijk}\epsilon^{abc}E_{ai}E_{bj}E_{ck}$
with each dreibein operator (as suggested in Ref.\cite{Thiemann})
expressed as $E_{ai} = \frac{1}{4{\pi}l^2_P\gamma} [V, A_{ai}]$.
Since the constraint Eq.(\ref{e:WD}) is now non-polynomial in the
basic conjugate variables, the quantum constraint does not
correspond to a simple factor ordering of the basic variables as
for the case of the polynomial constraint in Eq.(\ref{e:abf}).
Thus Eq.(\ref{e:WD}) discussed here is not motivated by simple
factor ordering. Although the classical Poisson bracket $\{{\tilde
v}, {\cal C}[A] \}_{P.B.} =
\pm\lambda(\frac{4\pi{G}}{ic^3}){\tilde v}$ is true, and may
appear as a motivation for Eq.(\ref{e:WD}), there is no rigorous
justification for promoting Poisson brackets between composite
operators to quantum commutator relations.

It may also be worth pointing out that the Chern-Simons operator
should have the characteristic property that for a 2-surface $S$
spanned by arbitrary $\epsilon_{abc}E^b\wedge E^c$, its commutator
with the basic area operator ${\cal A}_a \equiv \frac{1}{2}\int_S
\epsilon_{abc} E^b\wedge E^c$ should yield (in units of
$8\pi{l^2_P}$) the total non-Abelian magnetic flux traversing $S$
i.e. $[{\cal A}_a, {\cal C}] = (8\pi{l^2_P}{\gamma}i)\int_S F_a$.

\section{Further comments}

It is also interesting to investigate the situation for more
conventional phase space variables. If $(\tE, k)$ is adopted as
the fundamental pair, we may revert to the last equality in the
expansion of Eq.({\ref{e:kkk}}) for ${\cal C}[A=\gamma k
+\Gamma]$. However, a simplification occurs here in that ${\cal
C}[\Gamma]$ commutes with $\tv$ and $[\tv^2, \int_M k_a \wedge
(D^\Gamma k)^a]$ also vanishes if the Gauss Law constraint holds.
Furthermore the $\gamma^2$ (cubic in $k$) term of ${\cal C}[\gamma
k + \Gamma]/\gamma$ (please see Eq.({\ref{e:kkk}})) cancels the
$\gamma^2{\cal D}$ term in Eq.({\ref{e:abf}}). Thus the ADM
super-Hamiltonian constraint with $(\tE, k)$ as fundamental
variables may be simplified to \be [\tv^2, (\int_M
R_\Gamma^a\wedge k_a) + \frac{\lambda}{3}{\cal K} - {\cal D}]
=0;\en and we may use the last equality of Eq.({\ref{e:kkl}}) to
express ${\cal K}$ in terms of $(\tE, k)$. However, $R_\Gamma^a$
remains a non-polynomial function of $\tE$; but the formalism may
become simplify greatly in the minisuperpace context when
spatially flat slices can be chosen.

Since $[\tv^2, {\cal K}] \propto \tv^2$, there is a choice in
using only $\tv^2, {\cal C}$ (together with ${\cal D}$ for
($\gamma \neq \mp i$) in the reformulation if we do not insist on
expressing the constraint as a commutator. As indicated in the
later part of the Section III, this may be desirable in some
contexts. However if we wish to utilize $\cal K$, as alluded to
earlier in Section I earlier, it is possible to rewrite ${\cal K}$
in terms of $A$ and $V$ through Eq.({\ref{e:pc}}), yielding ${\cal
K} \propto \int_M \ept^{ijk}[A^a_{i}, V](D^A_j[A_{k}, V])_a\,
d^3x$.

A choice of $\gamma$ other than the self-dual or antiself-dual
values $\mp i$ appears to introduce formidable complications in
the commutator, regardless of whether $(\tE, A)$ or $(\tE, k)$ is
used as the basic conjugate pair. In the former instance, we
should either express $k_{ai} = (A_{ai} -\Gamma_{ai})/\gamma$ with
non-polynomial $\tE$ dependence in $\Gamma$ in ${\cal D}$; or
overcome the non-polynomial nature by writing $k_{ai} \propto
[A_{ai}, \cal K]$ with ${\cal K}$ expressed as in the previous
paragraph, or resort to an alternative expression for this
composite operator wherein the loop representation of the operator
${\cal K} \propto [V, \int {\tilde B}^{ia}[A_{ia}, V]d^3x]$ has
also been addressed in detail in Refs.\cite{Thiemann,Borissov}.

As it is well known, both the intrinsic and extrinsic curvatures
contribute to the Wheeler-DeWitt constraint. In the ADM
formulation the intrinsic curvature is related to the metric,
while its conjugate momentum is associated with the extrinsic
curvature. For the reformulated equation, $\tv$ can be related
only to the intrinsic curvature, but ${\cal C}[A = {\gamma}k +
\Gamma]$ is a carrier of information of both types of curvature.

\section*{Acknowledgments} The research for this work has been
supported in part by funds from the National Science Council of
Taiwan under Grant Nos. NSC94-2112-M-006-006 and
NSC95-2112-M-006-011-MY3. In expressing the super-Hamiltonian
constraint as a vanishing commmutator, the extension to consider
arbitrary values of the Immirzi parameter arose from a
correspondence from James B. Bjorken.

\end{document}